*Constructional details for*

# A simple atmospheric electrical instrument for educational use

A. J. Bennett[1] and R.G. Harrison

**Abstract**
Electricity in the atmosphere provides an ideal topic for educational outreach in environmental science. To support this objective, a simple instrument to measure real atmospheric electrical parameters has been developed and its performance evaluated. This project compliments educational activities undertaken by the Coupling of Atmospheric Layers (CAL) European research collaboration. The new instrument is inexpensive to construct and simple to operate, readily allowing it to be used in schools as well as at the undergraduate University level. It is suited to students at a variety of different educational levels, as the results can be analysed with different levels of sophistication. Students can make measurements of the fair weather electric field and current density, thereby gaining an understanding of the electrical nature of the atmosphere. This work was stimulated by the centenary of the 1906 paper in which C.T.R. Wilson described a new apparatus to measure the electric field and conduction current density. Measurements using instruments based on the same principles continued regularly in the UK until 1979. The instrument proposed is based on the same physical principles as C.T.R. Wilson's 1906 instrument.

**Keywords**: electrostatics; potential gradient; air-earth current density; meteorology;

## *Overview*

This instrument consists of a printed electrode with defined geometry, to which an ultra high input impedance unit gain voltage follower is connected. This provides the voltage output. The output voltage is also applied to a printed guard electrode, to minimise leakage.

## *Assembly*

### Making the PCB

The PCB is made from double sided copper clad PCB material. A complete sheet of 160mm x 100mm board is used. One side (the "bottom" side) of the PCB has the circuit tracks and the detector plate. The top side is the earth plate with the copper removed from the component area. To make the board proceed as follows:
- Print out the mask and carefully crop the image with approximately 10mm spare on all sides. Now draw an accurate line with a pen 5mm from the image on the right and bottom sides. This is a guide for positioning the board so that both sides etch correctly.
- First expose the top side in the UV box. First remove the protective covering from the PCB sheet where the copper is to be removed. This is done by carefully marking out the area and cutting away just the component area. This area is directly exposed in the light box.
- Expose the track side by positioning the PCB sheet on the marked out image. Remove the covering and expose the track side.
- Etching and tinning are performed as normal, ensuring both sides of the board are processed.

---

[1] *Department of Meteorology, The University of Reading. P.O. Box 243, Earley Gate, Reading RG6 6BB, UK. E-mail: a.j.bennett@reading.ac.uk*

## Components

Assemble board as normal until IC1 and R1, which need to be air wired. IC2 is mounted in a socket except for pin 3, which is bent out horizontally. Drill a 3mm hole to the left of IC2 pin 3 (the board is marked out for it). Solder a piece of tinned copper wire to the detector plate (track side). The wire goes through the hole. R1 is soldered between the electrode wire and pin 3 of IC1.

## Connections

- Earth plate

To connect the earth of the circuit to the earth back plate, drill a 1mm hole in the earth area at the bottom of the circuit. The hole in marked out on the board. Solder a piece of tinned copper wire into this hole. On the earth side bend the wire over the earth plate and solder it.

*Summary image: (viewed from bottom side)*

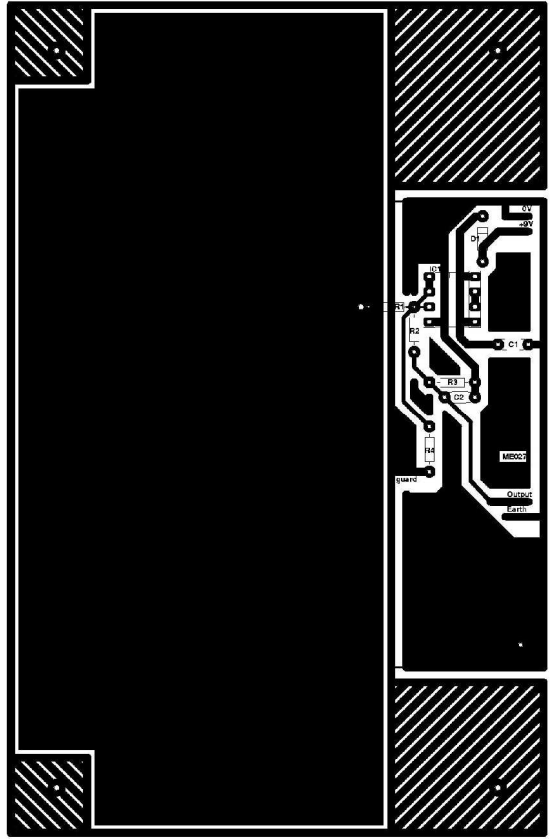

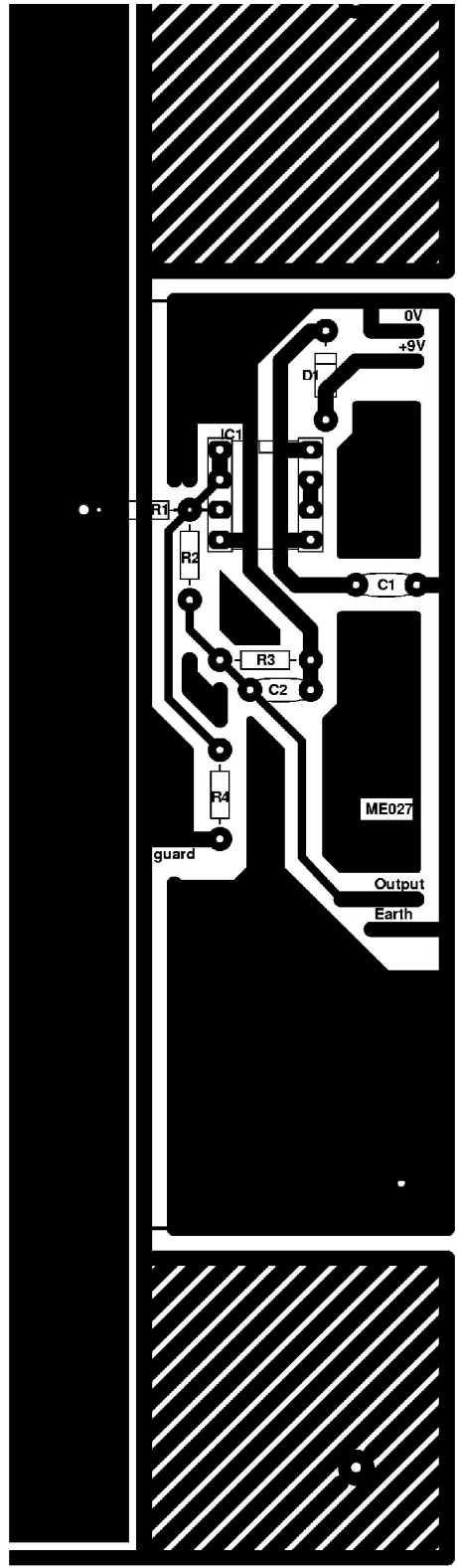

Scale: Expanded to show the component area

## *PCB (track layer)*

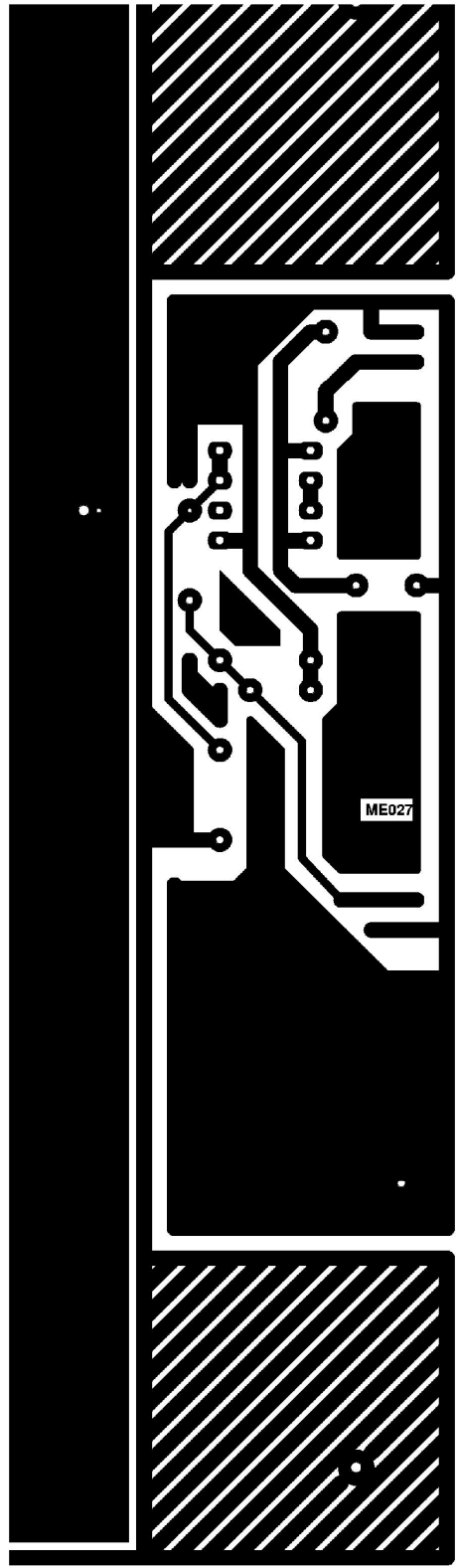

Scale: Expanded to show the component area

## *PCB (components/links layer)*

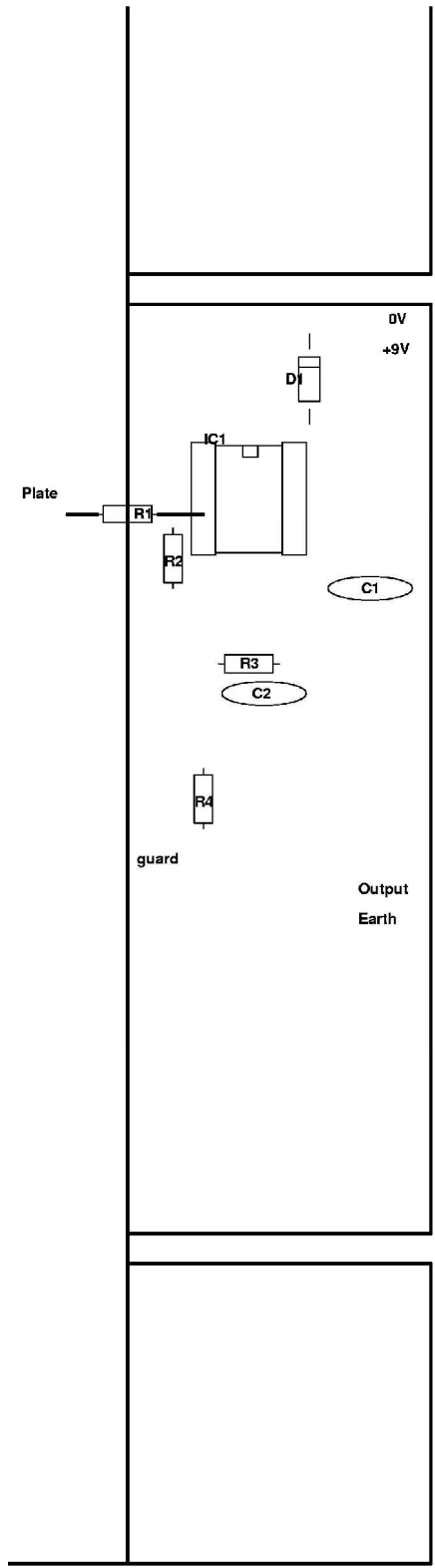

Scale: Expanded to show the component area

*Image of board*

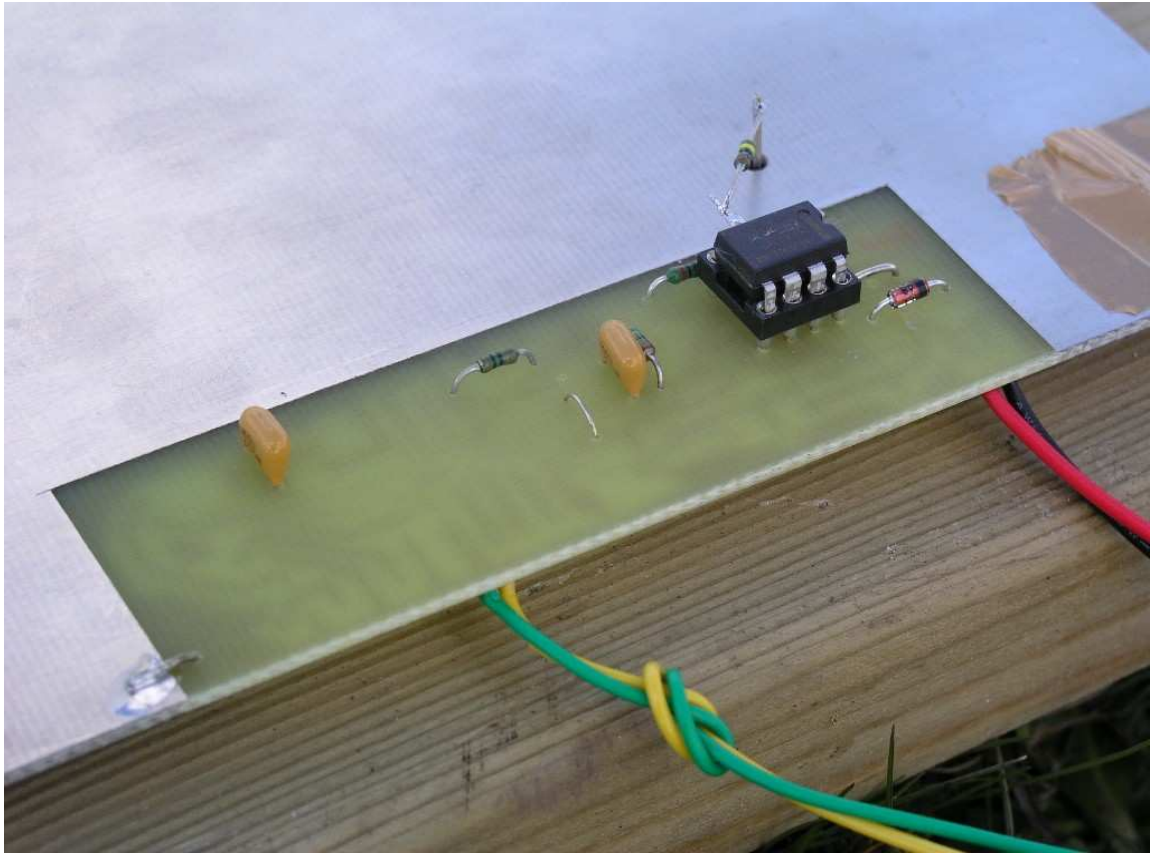

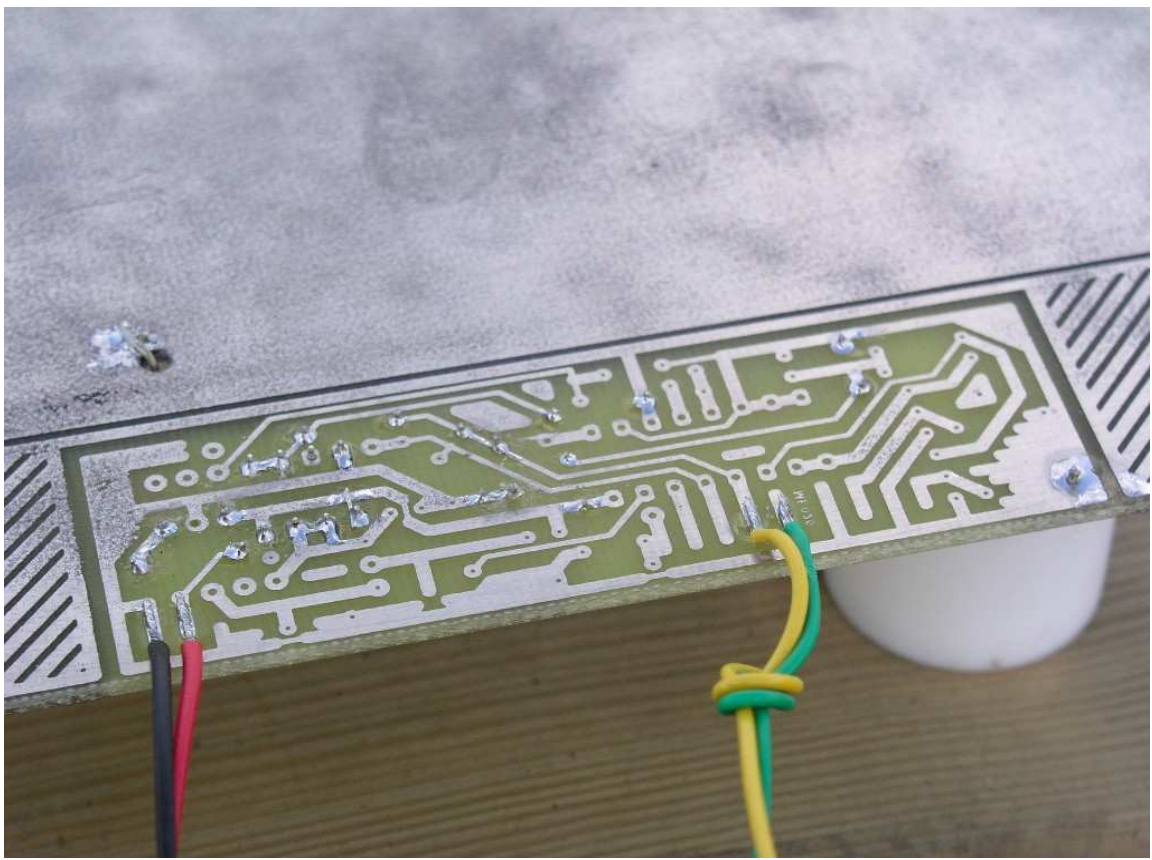

## Circuit schematic

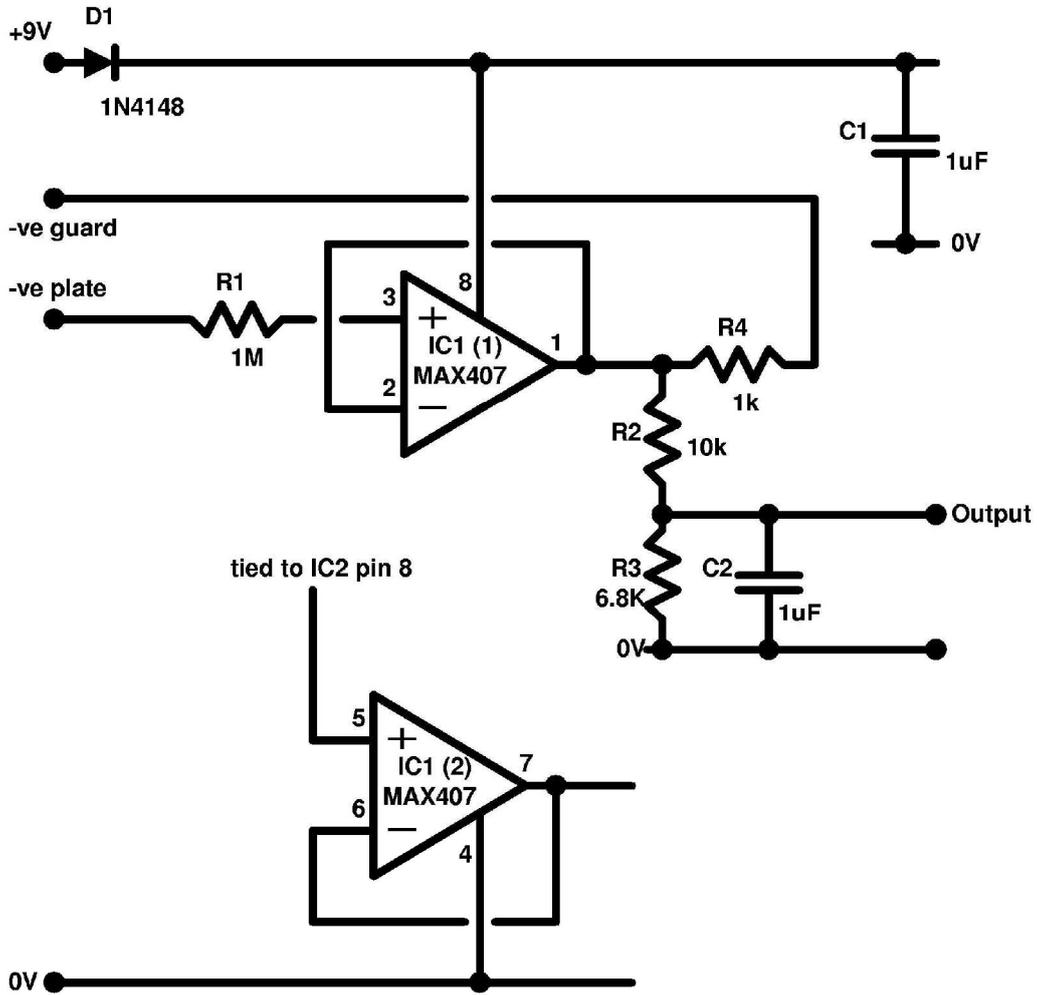

## *Component list*

| Ref | Value | Description |
| --- | --- | --- |
| R1 | 1M | 1/8 W metal film resistor, 2% |
| R2 | 10K | 1/8 W metal film resistor, 2% |
| R3 | 6K8 | 1/8 W metal film resistor, 2% |
| R4 | 1K | 1/8 W metal film resistor, 2% |
| C1 | 1uF | Ceramic, 50V |
| C2 | 1uF | Ceramic, 50V |
| D1 | 1N4148 | Small signal diode |
| IC1 | MAX407 | Dual electrometer op-amp |
|  |  |  |
|  |  |  |